\documentclass{nature}

\usepackage{graphicx}
\usepackage{dcolumn}
\usepackage{bm}
\usepackage{comment}

\usepackage{hyperref}
\usepackage{xspace}
\usepackage{lineno}

\usepackage{color}

\makeatletter
\let\saved@includegraphics\includegraphics
\AtBeginDocument{\let\includegraphics\saved@includegraphics}

\makeatother

\title{Small coverage effect in epidemic network models shows that masks can become more effective with less people wearing them}
\author{Peter Klimek$^{1,2,*}$, Katharina Ledebur$^{1,2}$, Stefan Thurner$^{1,2,3}$}

\begin{document}

\maketitle

\begin{affiliations}
\item Section for Science of Complex Systems, Medical University of Vienna, Spitalgasse 23, A-1090 Vienna, Austria
\item Complexity Science Hub Vienna, Josefst\"adterstra\ss e 39, A-1080 Vienna, Austria
\item Santa Fe Institute, 1399 Hyde Park Road, Santa Fe, NM 87501, USA
\end{affiliations}

$^*$Correspondence: peter.klimek@meduniwien.ac.at

\begin{abstract}

The effectiveness of non-pharmaceutical interventions to curb the spread of SARS-CoV-2 is determined by numerous contextual factors, including adherence.
Conventional wisdom holds that the effectiveness of protective behaviour such as wearing masks always increases with the number of people adopting it.
Here we show in a simulation study that this is not true in general.
We employ a parsimonious network model based on the well-established empirical facts that (i) adherence to such interventions wanes over time and (ii) individuals tend to align their adoption strategies with their close social ties (homophily).
When combining these assumptions, a broad dynamical regime emerges where the individual-level infection risk reduction for those adopting protective behaviour increases as the adherence to protective behavior decreases.
For instance, for a protective coverage of 10\% we find the infection risk for adopting individuals can be reduced by close to 30\% compared to situations where the coverage is 60\%.
Using estimates for the effectiveness of surgical masks, we find that reductions in relative risk of masking versus non-masking individuals range between 5\% and 15\%, i.e., vary by a factor of three.
This {\em small coverage effect} originates from system-dynamical network properties that conspire to increase the chance that an outbreak will be over before the pathogen is able to invade small but tightly connected  groups of individuals that protect themselves.
Our results contradict the popular belief that masking becomes ineffectual as more people drop their masks and might have far-reaching implications for the protection of vulnerable population groups under resurgent infection waves.
\end{abstract}

\maketitle

\section{Introduction}

The SARS-CoV-2 pandemic led to the widespread and sustained implementation of a number of non-pharmaceutical interventions (NPIs) such as social distancing, wearing face masks, or test-trace-isolate strategies. \cite{cheng2020covid,hale2021global,desvars2020structured}
From a scientific viewpoint, this opened the unique opportunity to accumulate real-world evidence on the effectiveness of 
such 
NPIs.\cite{flaxman2020estimating,haug2020ranking,sharma2021understanding, liu2021impact}
Meanwhile, there is  overwhelming empirical evidence that many of these interventions do indeed curb the virus spread. \cite{mendez2021systematic, chu2020physical}
The literature is less unanimous when it comes to concise quantitative estimates of the effectiveness of NPIs. \cite{haug2020ranking,liu2021impact,mendez2021systematic,talic2021effectiveness}
The heterogeneity in these estimates can be explained by a plethora of contextual factors impacting their adoption, ranging from geographic, cultural, socio-economic and healthcare-related to behavioural factors. \cite{pluemper2020covid,buja2020demographic,banholzer2022estimating, ledebur2022meteorological}

It has been shown that the extent to which individuals adhere to an NPI changes substantially over time even if no formal change in strengthening or easing the NPIs has occurred.\cite{petherick2021worldwide}
Adherence to physical distancing measures has been observed to wane over time. \cite{petherick2021worldwide, crane2021change, goldstein2021lockdown, reisch2021behavioral}
Adherence to low-cost habituating measures like mask wearing increased during the early phases of the pandemic, likely linked to extensive public health messaging. \cite{smith2022engagement, crane2021change, clinton2021changes}
However, even for those low-cost measures adherence plateaued and started to decline after vaccines were rolled out, the Omicron variant became dominant and an increasing number of countries dropped mask mandates. \cite{tjaden2022association, peixoto2022really, apnorc2022worries, office2022public}

Further, adherence to NPIs is strongly related to homophily. \cite{mcpherson2001birds}
It is a well-established fact that individual behaviour and attitudes cluster in social networks, meaning that attitudes of a person often conform with the attitudes of its close social contacts. \cite{newman2002assortative, bond201261}
Homophily is observed in health behaviour in general \cite{smith2008social,centola2011experimental} and more recently was confirmed in the context of the COVID-19 pandemic. \cite{bavel2020using, agranov2021importance} 
For instance, the likelihood that an individual wears a mask is closely related to the frequency of actual mask wearing, both in the general population, and in close social contacts. \cite{woodcock2021role,shin2022mask}

In this work by means of a simulation study we show that an additional factor may modulate the effectiveness of a wide range of NPIs such as wearing face masks.
We consider the broad class of NPIs that may --or may not-- be adopted by individuals and that reduce the chance of a transmission whenever 
adopted by either the transmitting or receiving individual.
Obviously, face masks are of this type but also other habituation measures such as keeping distance can reduce the chance of transmission, independent from whether the adopting individual is the transmitter or receiver.

In epidemiological models it is well-established that if the rate of adoption (coverage rate) and the effectiveness of NPIs are high enough, epidemic outbreaks can be successfully suppressed.
Consequently, the effectiveness of an NPI on the individual level (i.e., how much wearing a mask reduces one's personal risk of becoming infected) depends on system-dynamical properties such as the fraction of the entire population adopting the NPI.
As a higher coverage rate  moves the system closer to its epidemic threshold, higher population-level adoption typically also means a stronger individual-level infection risk reduction.


Here we aim to understand how the individual- and population-level effectiveness of mask-wearing depends on the coverage rate.
To put it more colloquially, if my own benefit from wearing a mask increases with the number of others wearing a mask, should I wear one, if no-one else is wearing one?
We study this question in a simple epidemiological model that acknowledges two repeatedly confirmed properties regarding the adoption of NPIs that were observed throughout the pandemic.
First, we assume that NPI adherence wanes over time;
someone who initially wore a mask during almost all non-household contacts will, over time, wear the mask at fewer and fewer occasions.
Second, we assume a certain degree of behavioural homophily, in the sense that the adoption behavior of NPIs clusters in social networks: individuals with close
social ties tend to synchronize their individual decisions on protection measures.
In all other aspects, we opt for a parsimonious modelling approach so that we can clearly isolate the effects originating from these assumptions and how they play out on the relation between individual- and system-level effectiveness of NPIs.

\begin{figure*}[htbp]
    \centering
    \includegraphics[width=0.6\textwidth]{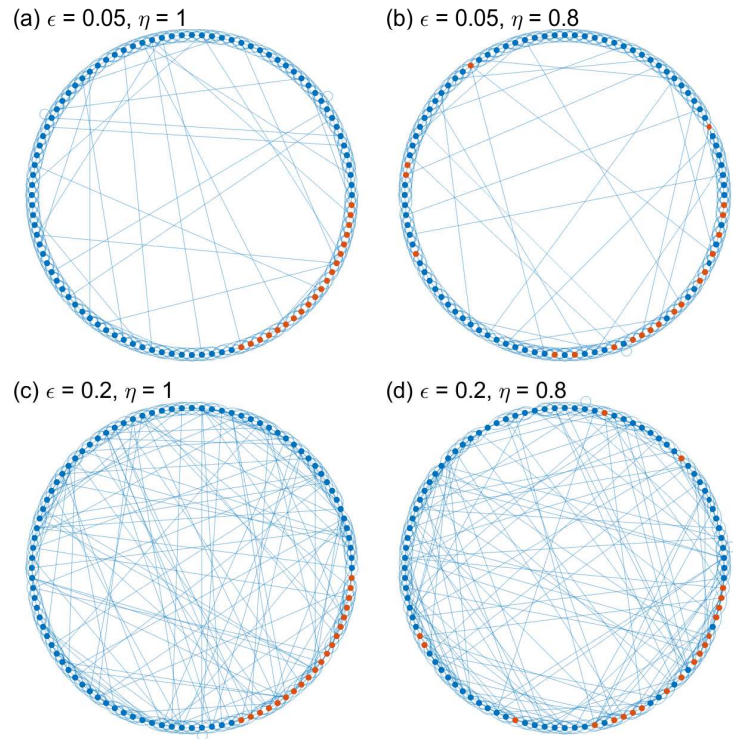}
    \caption{Illustration of a small-world network with assortative adoption of protective behaviour (homophily). We show $N=100$ individuals (nodes) out of which $20$ adopt (red) and $80$ do not adopt the behaviou (blue). Nodes are positioned on a ring and linked to their $k=6$ nearest neighbours; individuals with the same behaviour cluster in the network. To reflect imperfect assortative mixing, with some probability ($\epsilon$) a link is chosen and rewired toward a randomly chosen different node and with another probability (parameter $\eta$) the behaviour state (colour) is swapped between two randomly chosen nodes. The resulting networks are illustrated for ($\epsilon$,$\eta$) being (a) ($0.05$, $1$), (b) ($0.05$, $0.8$), (c) ($0.2$, $1$), and (d) ($0.8$, $0.2$).}
   \label{fig:networks}
\end{figure*}

In the model, all individuals have a static state indicating whether they adopt protective behaviour or not.
The effectiveness by which the adoption of the behaviour reduces the probability for a successful transmission decreases over time to implement waning of NPI adherence.
Each individual is either susceptible, infected or recovered. \cite{anderson1992infectious}
To include homophily, epidemic spreading takes place on a social contact network that shows assortative mixing with respect to the adoption of protective behaviour.
Assortative mixing is implemented with a small-world network model, see figure~\ref{fig:networks}. \cite{watts1998collective}
In this network, $N$ nodes are positioned on a ring and are linked to their $k$ nearest neighbors, resulting in a strongly clustered network, i.e., two neighbours of a given node are also likely to be connected.
With a certain probability, $\epsilon$, links are randomly rewired.
The parameter $\epsilon$ reflects a random linking between (family, work, and leisure) groups within the society. 
$\epsilon=0$ means a regular structure where everyone is linked to exactly $k$ neighbors; $\epsilon=1$ leads to a random network.
Finally, to introduce homophily in the adoption of protective behaviour, we first group those individuals (nodes) with the same behaviour on the ring and then randomly swap the behavior state between any two nodes with probability $1-\eta$, see figure~\ref{fig:networks}.
$\eta$ acts as a control parameter for the degree of homophily; large values represent a behaviorally clustered society, small values lead to a non-homophilic situation.

In the model, the paradigmatic SIR-type dynamics unfolds on networks of this type.
Initially, all nodes are in the susceptible state except for a small fraction of nodes in the infected state.
At each timestep a susceptible node can get infected from an infected neighboring node with probability, $\alpha$, if both nodes don't adopt protective behaviour.
Adoption of protective behaviour reduces the probability of a successful transmission by a time-dependent factor, $q_r(t)$, for the receiving individual and by $q_s(t)$ for the transmitter (source). 
We assume that $q_r(t)$ and $q_s(t)$ decrease at every timestep by a constant percentage to reflect waning of NPI adherence.
Nodes remain infected for $1/\beta$ timesteps, after which they become recovered and remain in that state for the rest of the simulation.

We measure how likely adopting and non-adopting individuals get infected during an outbreak.
We demonstrate that individual- and population-level infection risk depend on the coverage rate in a non-trivial way.

\section{Methods}

We consider a small world network of the Watts-Strogatz type with $N$ nodes positioned on a ring where they are connected to their $k$ nearest neighbours. 
With probability, $\epsilon$, a link is chosen, disconnected from one of its randomly chosen end nodes and rewired to any randomly chosen other node of the network (self-links are excluded).
The resulting network can be described by an $N \times N$ adjacency matrix, $A$, with entries, $A_{ij} = 1$, if nodes $i$ and $j$ are connected (neighboring) and $A_{ij} = 0$, otherwise.
Every node $i$ has a static state, $M_i$, signaling whether it adopts protective behaviour ($M_i=1$) or not ($M_i=0$) at the beginning of the simulation.
The coverage rate, $m$, represents the probability that a node adopts protective behaviour, $\sum_i M_i = mN$.
The {\em homophily parameter}, $\eta$, gives the fraction of nodes that do assortatively mix with each other.
That is, a number of $\eta m N$ nodes $i$, with $M_i=1$, are grouped next to each other on the network, whereas a fraction of $(1-\eta)mN$ of nodes with $M_i=1$ is randomly distributed across the ring, see figure~\ref{fig:networks}.

Every node, $i$, is endowed with a dynamic {\em epidemic state} variable, $X_i(t)$, that takes values in $X_i(t) \in \{S, I, R\}$ corresponding to the states in the SIR model: susceptible, infected, or recovered.
At every timestep, a susceptible individual $i$ can be infected with probability $\alpha$ from each of its neighbouring infected individuals $j$ if none of the two adopts protective behaviour ($M_i=M_j=0$).
If $M_i=1$, the infection risk decreases by a factor of $1-q_r(t)$, if $M_j=1$, the risk decreases by $1-q_s(t)$.
Adherence to these behaviours decreases exponentially with a rate $0 < q\leq 1$, such that $q_r(t) = q_r(0) q^t$ and $q_s(t) = q_s(0) q^t$, respectively.
Infected individuals recover with probability $\beta$.

At every timestep, $t$, the model loops over all nodes $i=1, \dots, N$ and performs a state update according to the following protocol (parallel update).
\begin{itemize}
    \item If $X_i(t)=S$, identify the set of all close infectious contacts of $i$, $Nb(i)$, as $Nb(i) = \{j|A_{ij}=1 \wedge X_j(t) = I \}$. For each node $j \in Nb(i)$, set $X_i(t+1)=I$ with probability, $\alpha\cdot(1-q_r(t))^{M_i}\cdot(1-q_s(t))^{M_j}$.
    \item If $X_i(t)=I$, set $X_i(t+1) = R$ with probability $\beta$.
    \item Proceed with the next node. Once all nodes are updated at time $t$, proceed to the next timestep, $t+1$.    
\end{itemize}
We denote the number of new cases by $C(t,M=1)$ and $C(t,M=0)$, for those that do and do not adopt protective behaviour, respectively.

We measure the risk of becoming infected during the outbreak on a individual and on a population scale.
While the individual infection risk gives the probability for an individual becoming infected as a function of whether protective behaviour is adopted or not, on the population-level infection risk measures the total number of infections in the entire population.
The individual infection risks in the two groups are $IIR(M=1) = \frac{\sum_t C(t,M=1)}{mN}$ and $IIR(M=0) = \frac{\sum_t C(t,M=0)}{(1-m)N}$, respectively; the population infection risk is given by $PIR=\frac{\sum_{t,M}C(t,M)}{N}$.

Unless stated otherwise, the following parameter settings were used.
Simulations were performed for $N=10^5$ nodes and averaged over 20 independent iterations.
Parameters have been chosen such that $k=20$, $\alpha=0.02$, $\beta = 0.1$, $q=0.99$, $q_r(0)=q_s(0)=1$, $\eta=1$, and $\epsilon=0.1$.
For the initial condition we randomly choose ten nodes and set their $X_i(t=0)=I$; for all the others we set $X_i(t=0) = S$. 
The model halts once the outbreak has ended, i.e., $X_i(t) \neq I$ for all $i$.

\section{Results}

\begin{figure*}[htbp]
    \centering
    \includegraphics[width=0.9\textwidth]{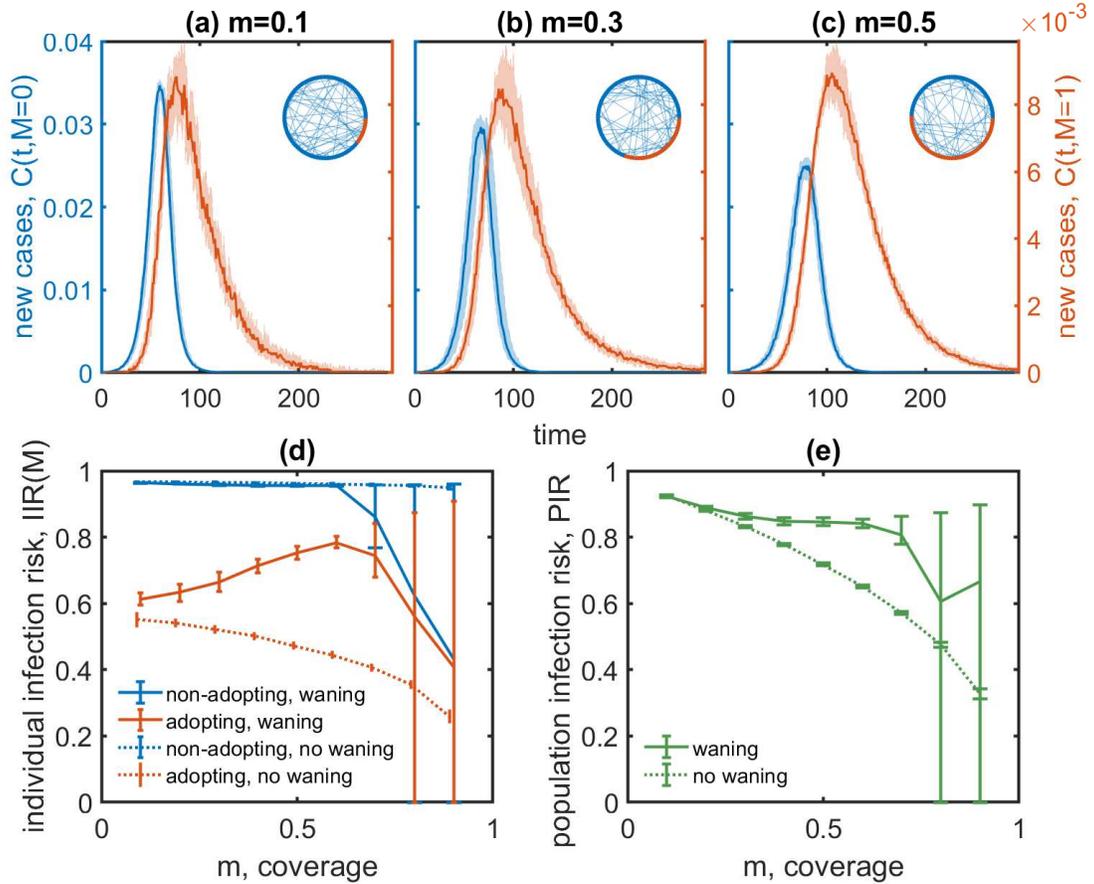}
    \caption{
    Demonstration of the small coverage effect of protective behaviour. The number of new cases over time for those adopting protective behaviour ($C(t,M=1)$, red) and those who do not ($C(t,M=0)$, blue) are shown for coverage from (a) $m=0.1$ over (b) $m=0.3$ to (c) $m=0.5$; shaded areas denote the 68\% CI. Insets illustrate the network and node states, $M$, for these settings, respectively. With increasing coverage, duration of the outbreak in adopting individuals increases. Results for the (d) individual infection risk,  $IIR(M=0,1)$, as a function of coverage $m$ are shown as solid lines and compared to a simulation for which mask-wearing adherence does not wane (dotted lines, using 
    $q=1$ and $q_r(0)=q_s(0)=0.4$); error bars denote the 68\% CI. If 
    adherence wanes, the infection risk has a clear maximum around $m=0.6$, while the infection risk is always a monotonously decreasing function of $m$ if adherence does not wane. On the (e) population level we observe a plateau with similar population infection risks, $PIR$, for a wide range of intermediate coverage.}
   \label{fig:runs}
\end{figure*}

Example runs for different coverage are shown in figure~\ref{fig:runs}(a)--(c).
We show the averages and the 68\% confidence intervals (CI) of the number of new cases as a function of time over multiple iterations for increasing coverage, $m$, i.e., increasing adoption of protective behaviour.
For individuals $i$ that do not adopt protective behaviour ($M_i=0$), the height of the epidemic peaks decreases with increasing coverage as the source-controlling effects indirectly protect them.
The duration of the outbreak increases however, leading to a situation where the total infection risk for non-adopting individuals during the outbreak changes little for sufficiently low levels of $m$, see the corresponding plateau for low $m$ in figure~\ref{fig:runs}(d).

Intriguingly, for adopting individuals the individual infection risk, $IIR$ may exhibit a pronounced maximum at an intermediate coverage rate.
Simulation runs for different values of coverage ($m$) show comparable heights of the epidemic peak for adopting individuals; see figure~\ref{fig:runs}(a)--(c).
However, the duration of this peak increases substantially with increasing $m$.
As a result, there is a regime where the individual infection risk for adopting individuals decreases with decreasing coverage, as shown in figure~\ref{fig:runs}(d).
For the settings used, 61\% of the adopting individuals get infected during the outbreak if they make up only 10\% of the population, but 78\% get infected if they account for 60\% of the total population.
We refer to this phenomenon 
as the ``small coverage effect''.

On a population level, this small coverage effect reveals itself in the form of a plateau in the population infection risk, $PIR$ at intermediate coverage, see figure~\ref{fig:runs}(e).
On this plateau the $PIR$-increasing small coverage effect approximately cancels with the increase in the number of adopting individuals with higher $IIR$.
For reference, figure~\ref{fig:runs}(e) also shows results for the $PIR$ without waning in which the monotonous decrease in $PIR$ under increasing coverage is recovered.

Different mechanisms are at work that reduce the individual infection risk for adopting individuals when coverage becomes very high or low, respectively.
Starting from a coverage of $80\%$ or more, simulation runs start to occur in which the epidemic dies out before adherence has waned sufficiently to trigger a resurgence of cases.
This can be seen in figure~\ref{fig:runs}(d) in the increase of the variance (confidence intervals) of the infection risk, $IR(M)$, for high coverage.
In this regime, the epidemic either quickly dies out (leading to values of $IR(M)$ close to zero) or larger outbreaks occur in both population groups if a small number of infections persists until adherence has sufficiently decreased.
One can think of this regime as an "all or nothing" scenario-driven by finite size effects in the network model.

In figure~\ref{fig:runs}(d,e) we also show simulation results for a scenario where adherence does not wane, i.e. $q=1$ (using $q_r(0)=q_s(0)=0.4$).
In this case, the frequently held belief of infection risk being a monotonously decreasing function of coverage, $m$, is recovered for both population groups and on an individual and population level.
That is, the more people adopt protective behaviour, the smaller the outbreak size.

Note that the small coverage effect is not a result of the normalization in the definition of the individual infection risks, $IIR(M=1)$ and $IIR(M=0)$.
As the infection risk for adopting individuals increases with $m$ for small coverage, figure~\ref{fig:runs}(d) shows that the number of infections grows even faster than the number of adopting individuals in that group.
The situation is slightly different on the population level, figure~\ref{fig:runs}(e), where this effect is compensated by an increasing number of individuals moving from the non-adopting to the adopting state and thereby reducing their infection risks.

\begin{figure*}[htbp]
    \centering
    \includegraphics[width=\textwidth]{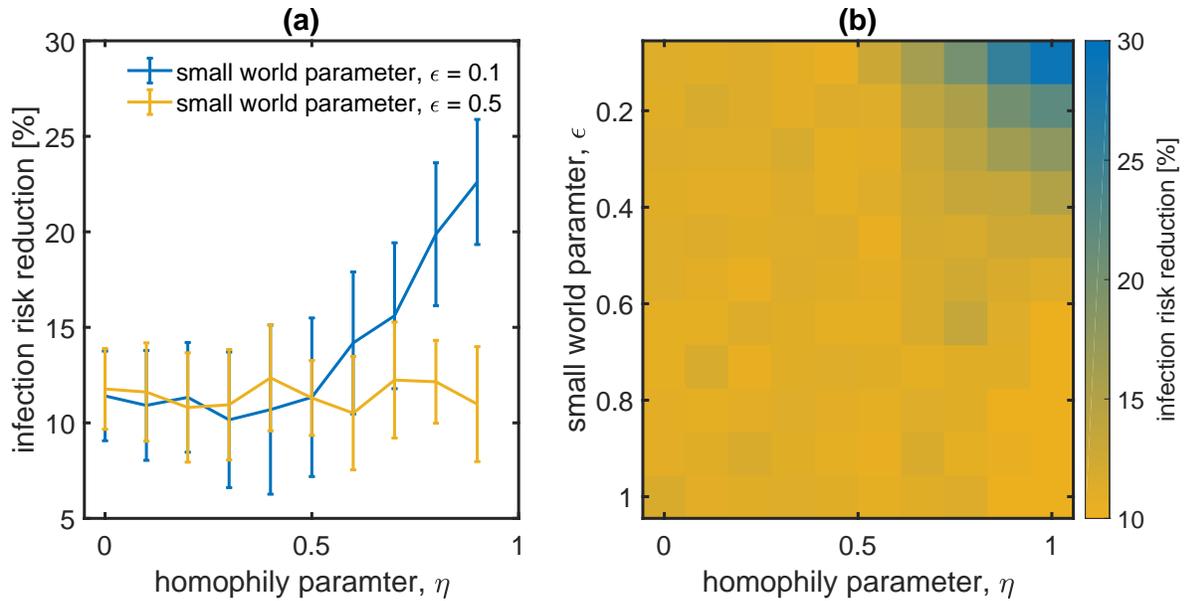}
    \caption{Homophily amplifies the small coverage effect. We show the reduction of the individual infection risk [\%] due to the small coverage effect from a coverage of of 60\% to 10\%. We find (a) an infection risk reduction of about $20\%-25\%$ for small world parameters representing strong homophily (large $\eta$). The reduction decreases with decreasing homophily and plateaus at values about $10\%$, for $\eta \leq 0.5$. Increasing the small world parameter from $\epsilon=0.1$ (blue) to $0.5$ (orange) reduces the small coverage effect even for high levels of homophily. (b) A parameter sweep over the homophily and small world parameters, $\eta$ and $\epsilon$, respectively, shows that the small coverage effect is most pronounced (close to 30\%) for small $\epsilon$ and large $\eta$ and plateaus between $10\%-15\%$ for other parameter settings.}
   \label{fig:homophily}
\end{figure*}

As seen in figure~\ref{fig:runs}(d), adopting behaviour at a coverage of 60\% ($m=0.6$) reduces the $IIR$ upon adopting protective behaviour from 95\% for $M=0$ to 78\% for $M=1$.
For a coverage of 10\% ($m=0.1$), however, adoption reduces the risk to 61\%.
To explore this effect more systematically for different parameter settings, we consider the infection risk reduction (in percent) from a coverage range of 60\% down to 10\%.
For the example given above, the small coverage effect thereby reduces the individual infection risk from 78\% for $m=0.6$ by 17\% (absolute reduction) to 61\% at $m=0.1$.

The extent to which homophily amplifies the small coverage effect is shown in figure~\ref{fig:homophily}.
Note that here we only show statistics from runs in which the epidemic does not ``die out'' immediately at the beginning of the simulation, as these runs would inflate the CIs.
For strong homophily ($\eta$ close to one), infection risk reduction falls within the range of 20\% to 25\%; see figure~\ref{fig:homophily}(a).
With $\eta$ decreasing, the infection risk reduction decreases too.
For values of around $\eta \approx 0.5$ the infection risk reduction bottoms out at 10\% where it assumes a plateau for weaker levels of homophily.
Increasing the small world parameter $\epsilon$ to 0.5 decreases assortative mixing and leads to a smaller infection risk reduction for large $\eta$.
These observations are corroborated by a parameter sweep over a range of values for $\eta$ and $\epsilon$; see figure~\ref{fig:homophily}(b).
Under strong assortative mixing (high $\eta$, small $\epsilon$), the small coverage effect reduces the infection risk by close to 30\%.
As assortative mixing decreases, the small coverage effect decreases and plateaus at around 10\% to 15\%.

So far we considered situations where the initial effectiveness of adopting protective behaviour is high ($q_r(0)=q_s(0)=1$).
Note that the small coverage effect can also be observed for less effective measures.
For instance, for cloth masks the inward and outward protective effectiveness has been estimated as 50\% and 30\%, respectively. \cite{pan2021inward} 
Using these values ($q_r(0)= 0.5$, $q_s(0)=0.3$), we observe infection risk reductions (from a coverage of 70\% to 10\%) of around $7\%$ (SD $1.6\%$).
One can also compare the relative risk ($RR$) of becoming infected between adopting and non-adopting individuals.
For a coverage of 70\%, adopting individuals had $RR=0.95$ compared to non-adopting individuals, for small coverage ($m=10\%$) we found $RR=0.88$. 
The magnitude of this reduction depends on the model parameters.
For instance, setting the connectivity of the network to $k=16$ gives a more pronounced infection risk reduction (at 10\% compared to a coverage 90\%) of $11\%$ (SD $2.1\%$), meaning that the relative risks of getting infected reduce from $RR=0.95$ to $0.84$.

\section{Discussion}

Conventional wisdom holds that the more people wear a mask, the higher is the protection it offers.
In this study we show that this is not necessarily true in general.
Using the plausible assumptions that (i) adherence to protective behaviour wanes over time and (ii) individuals tend to have a preference to interact more with like-minded others (homophily), we demonstrate the existence of a hitherto unknown small coverage effect:
As the number of people adopting the protective behaviour 
increases, the individual-level protection offered by the behaviour decreases.
This finding invalidates the frequently held notion that the individual-level effectiveness of measures like mask-wearing always increase with the number of people wearing a mask.\cite{stutt2020modelling, kai2020universal, catching2021examining, ngonghala2020mathematical,tian2021harnessing, howard2021evidence}
We numerically observed this effect for a wide range of parameters in a parsimonious network model.
The small coverage effect can be observed independently from how strong individuals cluster topologically in the network (i.e., the friend of a friend is likely to be a friend of mine too) and in terms of adopting the behaviour (homophily).
However, both types of clustering, when combined, do amplify the effect substantially.

The origin of the small coverage effect can be easily understood from system-dynamical properties.
If only a small fraction of the population adopts protective behaviour, there is a good chance that the outbreak will already be over in the non-adopting population group before measure adherence has completely waned.
The situation changes when there are more people that initially adopt the behaviour.
The overall susceptibility in the population will grow faster as the outbreak unfolds, as measure adherence is waning in a larger number of individuals.
This fuels--and thereby prolongs--the outbreak.
Hence, the chance that the outbreak is over before adherence has vanished completely decreases with coverage.

Homophily amplifies the small coverage effect.
This is likely related to local trapping of infection events in the network. \cite{allard2020role,thurner2020network}
Smaller groups of adopting individuals that preferentially have ties with each other are less likely to be affected in the initial phases of the outbreak, when adherence is still high.
As the outbreak continues, they are also less likely to be affected by large infection numbers amongst non-adopting individuals since they have a smaller number of ties with them.
Therefore, it will take the pathogen a longer to ``invade'' these protected groups, in contrast to a situation with no homophily.

It has indeed been observed that adherence to protective measures like mask wearing has strongly increased in the early phases of the pandemic \cite{smith2022engagement,crane2021change,clinton2021changes} but has since then decreased. \cite{apnorc2022worries,office2022public}
Many studies that tried to measure the real-world effectiveness of mask wearing basically compared the infection risk in adopting versus non-adopting individuals.\cite{alihsan2022efficacy}
Systematic reviews found that effectiveness can be quite heterogeneous across different settings and studies.\cite{talic2021effectiveness, chou2020masks}
The reasons for this heterogeneity are not yet fully understood.
The present work demonstrates that such study designs can be severely affected by the confounding influence of the population-wide mask coverage, which is typically not considered as a covariate.
Using conservative estimates for the protective effectiveness of cloth masks, we find that reductions in relative risk of masking versus non-masking individuals range between $5\%$ and $15\%$.
Further research is needed to understand whether empirically measured real-world differences in the effectiveness of mask-wearing could originate from the small coverage effect.

Academically, it would be challenging in a next step to demonstrate the mechanisms behind the small coverage effects also in the framework of game theory. 
One could think of a setting where adopters are cooperators and non-adopters are defectors; the payoff function (infection risks for adopters and non-adopters) would depend on the respective group sizes, and the waning effect would have to be implemented by a continuous shift towards a larger fraction of non-adopters over time. 
However, it is not clear from the outset what would constitute a round or interaction in which this game would be played.
A conceivable option could be a model with waning immunity (i.e., transitions from the recovered to the susceptible state) and seasonal driving of the transmission risk, $\alpha$. 
Individuals could observe their infection risks during one outbreak and possibly adjust their behaviour (state $M$) in the next seasonal outbreak depending on whether their past behaviour protected them or not.

Our simulation study is limited due to the use of a parsimonious model that makes idealized assumptions about several aspects, such as the infectious disease dynamics or the contact network topology.
This study design was a deliberate choice in order to isolate the mechanism driving the small coverage effect.
It remains to be seen to what extent this effect is observed in more sophisticated and calibrated epidemiological models and, above all, in real-world settings.

If it were true that wearing a mask offers less protection as less individuals mask themselves, it would be rational to give up on this behaviour as it won't offer protection anymore if everyone else drops his or her mask.
An important implication of our work is that quite the opposite might be true.
Masking, when adopted by a minority, might actually facilitate ``sitting out'' a wave without becoming infected.
This finding might have important consequences for the protection of vulnerable population groups that are at increased risk for severe disease.

{\bf References}
\bibliographystyle{ieeetr}  
\bibliography{references}

\end{document}